\documentclass[a4paper,11pt]{article}

\usepackage{jcappub}
\usepackage[utf8]{inputenc}

\usepackage{siunitx}

\usepackage{graphicx}
\usepackage{xcolor}

\usepackage{caption}
\usepackage{subcaption}
\usepackage{amsmath}
\usepackage{amssymb}
\usepackage{float}
\usepackage{threeparttable}

\usepackage{natbib}
\def\be{\begin{equation}}
\def\ee{\end{equation}}
\def\bea{\begin{eqnarray}}
\def\eea{\end{eqnarray}}

\def\cm{{\rm cm}}

\def\pc{{\rm pc}}
\def\kpc{{\rm kpc}}

\def\gev{{\rm GeV}}

\title{
The Velocity-Dependent $J$-factor of the Milky Way Halo:
Does What Happens in the Galactic Bulge Stay in the Galactic Bulge?
}
\author[a]{Kenny Kiriu,}
\author[a]{Jason Kumar,}
\author[a]{Jack Runburg}

\affiliation[a]{\mbox{Department of Physics \& Astronomy, University of Hawai`i, Honolulu, HI 96822, USA}}

\emailAdd{kiriuk@hawaii.edu}
\emailAdd{jkumar@hawaii.edu}
\emailAdd{runburg@hawaii.edu}

\abstract{We consider the angular distribution of the photon signal which could arise
from velocity-dependent dark matter annihilation within the Galactic bulge.  We find that, for the
case of Sommerfeld-enhanced annihilation, dark matter annihilation within the bulge is
dominated by slow speed particles which never leave the bulge, allowing one to find a simple
analytic relationship between the dark matter profile within the Galactic bulge and the angular
distribution.  On the other hand, for the case $p$- or $d$-wave annihilation, we find that
the small fraction of high-speed particles which can leave the bulge provide a significant,
often dominant, contribution to dark matter annihilation within the bulge.  For these scenarios,
fully understanding dark matter annihilation deep within the Galactic bulge, and the angular distribution of
the resulting photon signal, requires an understanding of
the dark matter profile well outside the bulge.  We consider the Galactic Center excess in
light of these results, and find that an explanation of this excess in terms of $p$-wave
annihilation would require the dark matter profile within the bulge to have a much steeper
profile than usually considered, but with uncertainties related to the behavior of the profile
outside the bulge.}

\begin{document}
\maketitle

\section{Introduction}

The Galactic Center (GC) of the Milky Way (MW) is an interesting target for the indirect
detection of dark matter, because it is expected to have a large density of
dark matter (DM).  In fact, an excess of gamma-rays in the GeV range has been observed
within the inner several degrees of the GC~\cite{Goodenough:2009gk,Hooper:2010mq,Fermi-LAT:2015sau}.  The origin of these photons is under study,
and may lie in ordinary astrophysical sources, such as millisecond pulsars (MSPs) (see, for example,~\cite{Abazajian:2010zy}).
Dark matter is another possible origin which has been the subject of much study, particularly
because the angular distribution seen in the photon excess is consistent with what one
would expect from dark matter annihilation, assuming a density distribution consistent
with results from numerical simulations.  In this work, we will study the dependence of
the gamma-ray angular distribution on the velocity-dependence of dark matter annihilation.
Although our analysis will be general, we will consider an application of these results
to the observed GC excess, assuming its origin is dark matter annihilation.

The velocity-dependence of the dark matter annihilation cross section impacts  the consistency of a dark matter explanation for the GC excess with constraints
from searches of dwarf spheroidal
galaxies (dSphs)~\cite{Fermi-LAT:2010cni,Fermi-LAT:2011vow,Fermi-LAT:2013sme,Fermi-LAT:2015att,Fermi-LAT:2016uux}.
In the most commonly studied scenario ($s$-wave
annihilation) $\sigma v$ is independent of the relative velocity $v$.  In this case, models
which can explain the GC excess have cross sections which are roughly at the limit of dSphs
searches.  Although there are systematic uncertainties which make a clear statement difficult,
it is fair to say that an explanation of the GC excess through $s$-wave dark matter annihilation
faces non-trivial constraints from dSphs searches~\cite{Fermi-LAT:2016uux,Chang:2018bpt,Hooper:2019xss}.  But if dark matter annihilates from a
$p$- or $d$-wave initial state, then annihilation is more heavily suppressed in regions where the
dark matter relative velocity is smaller, suppressing the annihilation rate in dSphs relative to
the GC.  Since these scenarios effectively weaken any constraints from dSphs searches on explanations
of the GC excess, it is important to know how these scenarios affect the effective $J$-factor, which
encodes the angular distribution of the signal.

There has already been previous work discussing the effective $J$-factor of halos in the
case of velocity-dependent dark matter annihilation (see, for
example,~\cite{Robertson_2009,Ferrer:2013cla,Boddy:2017vpe,Zhao:2017dln,Petac:2018gue,Lacroix:2018qqh,Boddy:2019wfg}),
and of the GC in particular~\cite{Boddy:2018ike,Johnson:2019hsm,Board:2021bwj,McKeown:2021sob}.
The case of
the GC is complicated by the fact that there is a large baryonic contribution to the
gravitational potential, yielding  additional parameters of the baryonic distribution
which affect the $J$-factor.  In this
work, we will consider dark matter annihilation within the Galactic bulge.  Under the approximation
that baryonic matter dominates the gravitational potential, which is taken to have a power-law expansion,
one can then solve for the dark matter velocity-distribution and the $J$-factor
analytically in the region of the Galactic bulge.

We will find that for Sommerfeld-enhanced dark matter annihilation, the angular distribution in bulge
region is dominated by the annihilation of low-speed particles which never leave the bulge, resulting in a
simple analytic prediction for the angular distribution which depends only on the slope of the dark matter profile inside the bulge.
This analytic prediction for the angular distribution can
be applied to a wide range of scenarios, including different choices for the dark matter density profile outside
the Galactic bulge.

On the other hand, we find that the $p$-wave and
$d$-wave annihilation rates receive a sizeable contribution from the energetic particles which explore the gravitational
potential far outside the bulge.  Although such particles only provide a negligible contribution
to the dark matter density inside the bulge, they are the fastest particles, and thus may
dominate the annihilation rate within the bulge.  In this case, one cannot disentangle the
angular distribution of the signal near the GC  from the gravitational potential
far from the GC.
As a result,
one will see deviations from the analytic prediction which depend on the steepness of the dark matter
profile, and find that a complete description of the
angular distribution at small angle requires knowledge of the dark matter distribution even well outside
the bulge.

It is generally difficult to determine the behavior of dark matter within the bulge region.  Stellar
data can by used to constrain the density distribution within the bulge, but there are large
uncertainties associated with the complexity of the stellar populations.  Results can also be
obtained from numerical simulations, but these simulations often lack the resolution to
adequately probe the bulge (for recent progress, see~\cite{McKeown:2021sob}).  Standard density profiles, such as NFW (or generalized versions
of NFW), typically assume a power-law behavior all the way out to the scale radius ($r_s \sim 21~\kpc$),
but there is no particular reason why there cannot be a different power-law behavior within
the bulge itself, where the baryonic potential is more important.  It is thus important to understand
the circumstances in which the photon angular distribution depend only on the dark matter profile near the
bulge, and how important corrections due to the behavior outside the bulge can be.

It has been found that, for $s$-wave annihilation, the GC excess morphology can be matched by
a dark matter density profile which scales as $\rho (r) \propto r^{-\gamma}$ with
$\gamma$ in the range $1.2 - 1.4$ (see, for example,~\cite{Hooper:2010mq,Calore:2014xka}).
For the case of $p$-wave annihilation, we find that the same morphology may require a steeper profile.
Even so, we find that an accurate model of the photon angular distribution within the bulge
requires knowledge of particles which exit the bulge.  This implies that ambiguities in the dark
matter profile and the effects of triaxiality, for example, can have a significant effect on
the angular distribution for $p$-wave annihilation in the GC region.

The plan of this paper is as follows.  In Section~\ref{sec:formalism}, we describe the
general formalism for computing the velocity-dependent $J$-factor in the Galactic bulge region.
In Section~\ref{sec:analytic}, we describe an analytic approximation to the angular distribution,
and the considerations which affect its validity.  In Section~\ref{sec:GCE}, we apply these results to the
GC excess.  We conclude with a discussion of our results in Section~\ref{sec:conclusion}.

\section{Dark matter and baryons near the galactic center}
\label{sec:formalism}

We consider the case in which the dark matter annihilation cross section has a
velocity dependence which can be expressed as
\bea
\sigma v &=& (\sigma v)_0 \times (v/c)^n ,
\eea
where $v$ is the relative speed 
and $(\sigma v)_0$ is a constant which is independent of $v$.  
The most common example is dark matter annihilation
from an $s$-wave initial state ($n=0$), in which case $\sigma v$ is independent of $v$ in the non-relativistic limit.  
But there are a variety of well-motivated scenarios which are worth considering.  If
dark matter annihilates from a $p$-wave initial state, then
one would find $n=2$; this scenario can arise, for example,
if dark matter is a Majorana fermion which couples to a
Standard Model (SM) fermions/anti-fermion pair through
interactions which respect minimal flavor violation (MFV)
(see, for example,~\cite{Kumar:2013iva}).  If the dominant
annihilation channel is from a $d$-wave initial state, then
one would find $n=4$; this scenario can arise if dark matter
were instead a self-conjugate spin-0 particle~\cite{Kumar:2013iva,Giacchino:2013bta,Toma:2013bka}.
If dark matter annihilation is Sommerfeld-enhanced due to
an attractive force mediated by a nearly massless particle,
one would find $n=-1$~\cite{Arkani-Hamed:2008hhe,Feng:2010zp}.

If we assume that the dark matter is a self-conjugate particle, we can express the photon flux arising from dark matter annihilation in the
Milky Way halo as
\bea
\frac{d^2 \Phi}{dE d\Omega} &=& \frac{(\sigma v)_0}{8\pi m_X^2}
\frac{dN}{dE} J_S (\cos \theta) ,
\eea
where $m_X$ is the dark matter mass, $dN / dE$ is the
photon energy spectrum arising from dark matter annihilation,
and
\bea
J_S (\cos \theta) &=& \int d\ell
\int d^3 v_1~f(r(\ell, \theta), v_1 )
\int d^3 v_2~f(r(\ell, \theta), v_2 )
\times (|\vec{v}_1 - \vec{v}_2|/c)^n .
\eea
Here, $f(r, v)$ is the dark matter velocity-distribution
within the halo, which we assume  is spherically symmetric
and isotropic.  This essentially implies that $f$ is a function only of
$r = |\vec{r}|$ and $v = |\vec{v}|$.
$D$ is the distance to the GC, $\theta$ is
the angle between the GC and the line-of-sight, and
$\ell = D\cos \theta \pm \sqrt{|\vec{r}|^2 - D^2 \sin^2 \theta}$
is the distance along the line-of-sight.

It will be convenient to express $J_S (\cos \theta)$ as
\bea
J_S (\cos \theta) &=& \int_0^\infty d\ell ~P_n^2 (r) ,
\nonumber\\
&\sim&  2 \int_{D \sin \theta}^\infty dr \left(1 - \frac{D^2}{r^2} \sin^2 \theta \right)^{-1/2}
P_n^2 (r) ,
\label{eq:JS_exact}
\eea
where
\bea
P_n^2 (r) &=& \int d^3 v_1 \int d^3 v_2~f(r, v_1 )~f(r, v_2 )
\times (|\vec{v}_1 - \vec{v}_2|/c)^n ,
\nonumber\\
&=& 8\pi^2 \int_0^\infty dv_1 \int_0^\infty dv_2~v_1^2 v_2^2~f(r, v_1 )~f(r, v_2 )
\frac{(v_1 + v_2)^{n+2} - |v_1-v_2|^{n+2}}{(n+2)v_1 v_2 c^n} .
\eea
Note that the upper limit integration in the second line of eq.~\ref{eq:JS_exact} encompass negative
values of $\ell$, including integration along the line-of-sight in both directions.
But when observing near the GC, the associated error is negligible.

To determine the $J$-factor, we need an expression for $f(r,v)$.
It follows from Liouville's Theorem that
the time-averaged velocity-distribution can only be a function of the energy, as
this is the only relevant integral of motion for a classical orbit.  Defining
$f(r,v) = f(E(r,v))$, where $E = (1/2) v^2 + \Phi(r)$ is the energy per mass of
a dark matter particle, and $\Phi (r)$ is the gravitational potential, we have
\bea
\rho (r) &=& 4\pi \int_0^{v_{esc}(r)} dv~v^2~f(r,v) ,
\nonumber\\
&=& 4\sqrt{2} \pi \int_{\Phi (r)}^{\Phi (\infty)} dE~\sqrt{E - \Phi(r)}~f(E) ,
\eea
where $v_{esc} (r)$ is the galactic escape velocity at $r$.  Inverting this
equation with the Abel integral equation yields the Eddington inversion formula
\bea
f(E) &=& \frac{1}{\sqrt{8} \pi^2} \int_E^{\Phi(\infty)} \frac{d^2 \rho}{d\Phi^2}
\frac{d\Phi}{\sqrt{\Phi - E}} ,
\eea
where we have implicitly expressed $\rho$ as a function of $\Phi$.  We may write
the gravitational potential $\Phi = \Phi_{DM} + \Phi_{bary}$ as the sum of the
potential due to dark matter and the potential due to baryonic matter in the Milky Way,
with
\bea
\Phi_{DM} (r) &=& \Phi_{DM} (0) +
4\pi G_N \int_0^r \frac{dx}{x^2} \int_0^x dy~y^2~\rho(y) .
\eea

We utilize a spherical approximation to the gravitational potential due to
baryonic matter in the bulge and
the disk~\cite{Strigari:2009zb,Pato:2012fw}, yielding
\bea
\Phi_{bary} (r) &=& -G_N \left[\frac{M_b}{c_0 +r} + \frac{M_d}{r} \left(1-e^{-r/b_d} \right)  \right]
+ G_N \left[\frac{M_b}{c_0} + \frac{M_d}{b_d} \right],
\eea
where we take $M_b = 1.5 \times 10^{10} M_\odot$ as the mass of the Galactic bulge and
$M_d = 7 \times 10^{10} M_\odot$ as the mass of the Galactic disk.  We take the bulge scale
radius to be $c_0 = 0.6~\kpc$, and the disk scale radius to be $b_d = 4~\kpc$.
We have added a convenient constant to the potential to set $\Phi(0)=0$.
Note, we
are ignoring the contribution to the gravitational potential due to the black hole at the
center of the Milky Way.  This contribution should be subleading for $r>\pc$ (see~\cite{Sandick:2016zeg},
for example).

Given an ansatz for $\rho(r)$, one can then numerically integrate the above equations to obtain
$J_S (\cos \theta)$~\cite{Boddy:2018ike}.  We will consider generalized NFW profiles, given by
\bea
\rho (r) &=& \frac{\rho_s}{(r/r_s)^\gamma (1+(r/r_s))^{3-\gamma}} ,
\eea
where $\gamma$ is a parameter describing the inner slope, $\rho_s$ is the scale density,
and $r_s$ is the scale radius, which we take to be $r_s = 21~\kpc$.

\section{Analytic Approximation}
\label{sec:analytic}

We will now consider an analytic approximation to $J_S$ at small $\theta$.  For this
purpose, we assume
\begin{itemize}
\item{The $J$-factor at small $\theta$ is dominated by dark matter annihilation within
the bulge, so we can ignore dark matter annihilation for $r>c_0$.}
\item{The dark matter density within the bulge can be written as
$\rho(r) \sim \rho_s (r/r_s)^{-\gamma}$.  This is true for the generalized NFW profiles
for which we obtained numerical results, but can encompass many more profiles.}
\item{The gravitational potential within the bulge is dominated by $\Phi_{bary}$, so we
can ignore $\Phi_{DM}$.}
\item{Within the bulge, $\Phi_{bary}(r)$ is sufficiently well-approximated by a Taylor expansion
to linear order in $r$.}
\end{itemize}
Note, we will not always find these assumptions to be valid.  As we will see in Section~\ref{subsec:validity},
some choices of the parameters will yield significant deviations from these assumptions, leading to
deviations from the analytic prediction.  We will address the import of deviations from the analytic treatment in
Sections ~\ref{subsec:validity} and~\ref{sec:GCE}.

But given these assumptions, we find
\bea
\Phi (r) \sim \Phi_0 r ,
\eea
where $\Phi_0 = G_N [(M_b / c_0^2) + M_d / 2b_d^2]$, given our spherical
approximation to the baryonic potential.
But we will see that the our result will apply
for any value of $\Phi_0$.  The only necessary condition is that the linear approximation
to the potential be sufficiently good within the bulge.

We then find
\bea
\rho (r)
&=&  4\sqrt{2} \pi \left(\Phi_0 r \right)^{3/2} \int_1^{\Phi(\infty) /\Phi_0 r} dx~
\sqrt{x-1}~f(x \Phi_0 r ) .
\label{eq:rho}
\eea
For $r \ll c_0$, we may take $\Phi(\infty) / \Phi_0 r \rightarrow \infty$, in which case
the only dependence of the
integral
on $r$ is in the argument
of $f$.
Since $\rho (r) \propto r^{-\gamma}$, we can solve eq.~\ref{eq:rho} by assuming a
power-law ansatz for $f$, namely
\bea
f(E) &=& f_0 E^{-\gamma -3/2} ,
\nonumber\\
f_0 &=& \rho_s (r_s\Phi_0)^{\gamma}  \left[4\sqrt{2} \pi \int_1^\infty dx~x^{-\gamma -3/2}\sqrt{x-1} \right]^{-1} .
\label{eq:f}
\eea
Note that the integral in eq.~\ref{eq:f} converges for $\gamma >0$.  In this case, the high-velocity
tail of particles which can leave the bulge contribute negligibly to the density at small $r$.

We now have
\bea
P_n^2
&=& 8\pi^2 f_0^2 (\Phi_0 r)^{-2\gamma +(n/2)} \int_0^\infty dy_1 \int_0^\infty dy_2~
 y_1^2 y_2^2~
\left((1/2)y_1^2 + 1 \right)^{-\gamma-3/2}
\left((1/2)y_2^2 + 1 \right)^{-\gamma-3/2}
\nonumber\\
&\,& \times
\frac{(y_1 + y_2)^{n+2} - |y_1-y_2|^{n+2}}{(n+2)y_1 y_2 c^n} ,
\nonumber\\
&=& \left( f_0 \Phi_0^{-\gamma} \right)^2 \left( \Phi_0/c^2 \right)^{ (n/2) }  I_{\gamma,n} r^{-2\gamma + (n/2)} ,
\eea
where
\bea
I_{\gamma,n} &\equiv& 8\pi^2\int_0^\infty dy_1 \int_0^\infty dy_2~
 y_1^2 y_2^2~
\left((1/2)y_1^2 + 1 \right)^{-\gamma-3/2}
\left((1/2)y_2^2 + 1 \right)^{-\gamma-3/2}
\nonumber\\
&\,& \times
\frac{(y_1 + y_2)^{n+2} - |y_1-y_2|^{n+2}}{(n+2)y_1 y_2 } .
\eea

Note, however, that the integral defining $I_{\gamma, n}$
only converges for $\gamma > n/2$.  For profiles which
are not steep enough, the divergence at large velocity
indicate that dark matter annihilation, even well within
the bulge, is dominated by high-velocity particles which
leave the bulge.  Although such particles make only a
negligible contribution to $\rho(r)$, the enhancement
to the annihilation cross section for high-velocity particles when
$n \geq 2\gamma$ means that they can dominate the
annihilation rate.

If $\gamma > n/2$, however, we can find an analytic
approximation to the $J$-factor of the GC within
the inner few degrees.  We will use our expression
for $J_S (\cos \theta)$ (eq.~\ref{eq:JS_exact}), but
only integrate to a distance $c_0$ from the GC, assuming
that this region dominates the annihilation rate.  This
expression is thus restricted to the angular region
$\theta \leq c_0 / D \sim 4^\circ$, for which we can
approximate $\sin \theta \sim \theta$.
We then find
\bea
J_S (\cos \theta) &=&
2 \left( f_0^2 \Phi_0^{-2\gamma + (n/2) }c^{-n} \right) I_{\gamma,n}
\int_{D\theta}^{c_0} dr
\left(1 - \frac{D^2}{r^2} \theta^2 \right)^{-1/2}
r^{-2\gamma + (n/2)} .
\eea
For $\theta \ll c_0 /D$, we can take the upper limit of
integration to infinity, in which case the integral has a
power-law dependence on $\theta$, yielding
\bea
J_S (\cos \theta) &\sim &
2D \left[ (f_0/(\Phi_0 D)^\gamma)^2 (\Phi_0 D /c^2)^{n/2}
I_{\gamma,n} \right] \left[
\int_1^{\infty} dx
\left(1 - x^{-2}  \right)^{-1/2} x^{-2\gamma + (n/2)} \right]
\nonumber\\
&\,& \times  \theta^{1-2\gamma + (n/2) } .
\eea
We thus see that, at small angles, there is a complete degeneracy between $\gamma$ and $n$, provided the
baryonic potential is dominant and $\gamma > n/2$.
For a sufficiently steep profile satisfying this condition,
the angular distribution of photon emission at small angles
is independent of the dark matter distribution outside the
bulge.  This condition is satisfied for all reasonable
models in the case of Sommerfeld-enhanced annihilation,
and for even moderately cuspy profiles in the case of
$s$-wave annihilation.  But it is only satisfied for profiles
cuspier than NFW in the case of $p$-wave annihilation, and
for very steep profiles ($n > 2$) in the case of $d$-wave
annihilation.

\subsection{Validity of the Analytic Approximation}
\label{subsec:validity}

\begin{figure}
    \centering
    \includegraphics{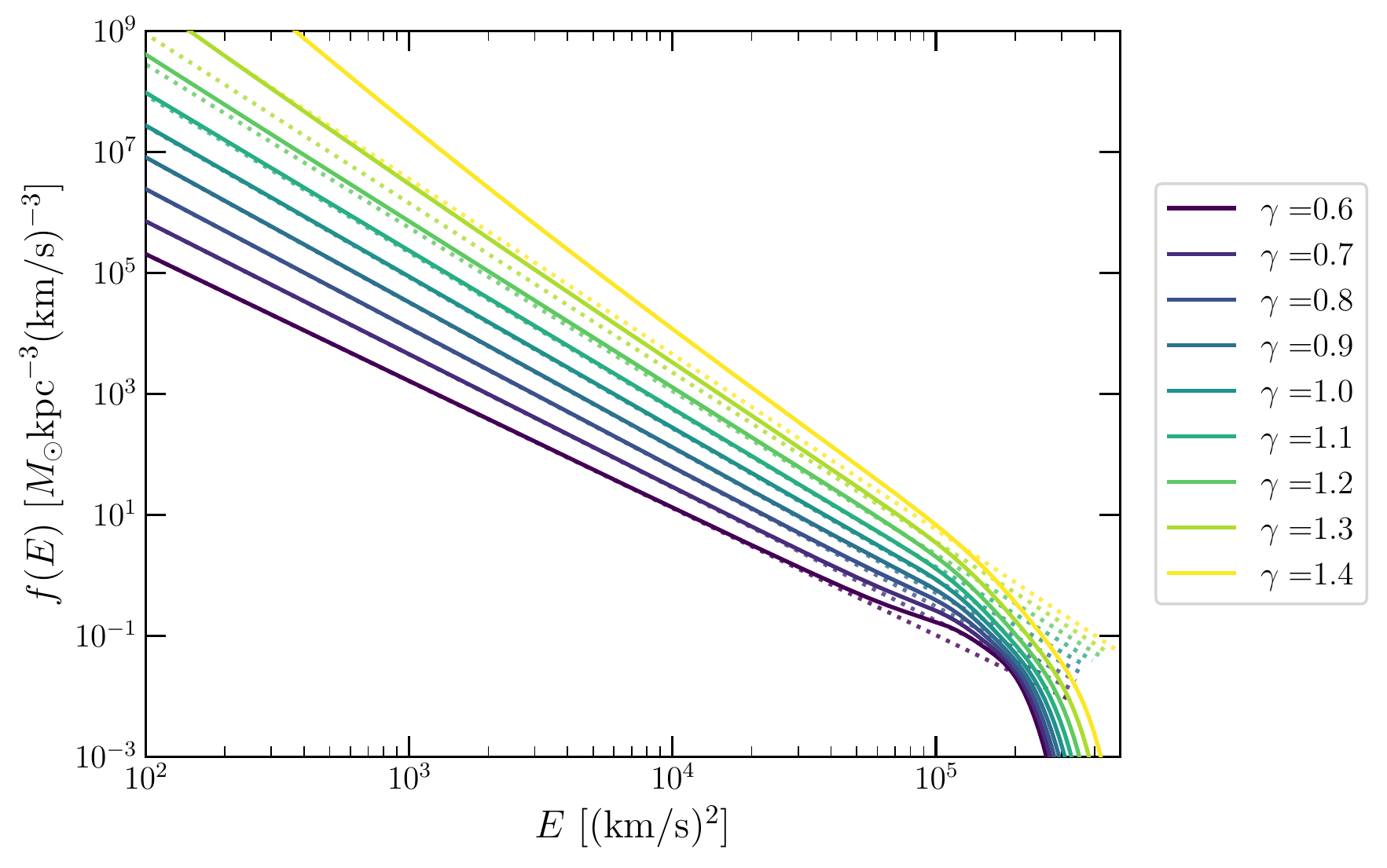}
    \caption{Plot of $f(E)$
    as a function of $E$,
    assuming  $\rho_s=8\times 10^6 M_\odot / \kpc^3$,
for various values of $\gamma$.  The solid lines show the complete numerical calculation, while the
dashed lines indicate the power-law analytic approximation.
}
    \label{fig:fe}
\end{figure}

There are several considerations which affect whether or not
this analytic approximation is valid.  Provided the baryonic
potential dominates over the dark matter gravitational potential,
the velocity-distribution at the core (that is, at small $E$)
is well-approximated by the analytic power-law form, as we
illustrate in Figure \ref{fig:fe}.
Here we plot $f(E)$
as a function of $E$,
taking
$\rho_s = 8 \times 10^6 M_\odot / \kpc^3$, for various choices of $\gamma$ (solid lines).  We also plot the
power-law analytic approximation (dashed lines).  We see that, for profiles which are
not too steep, the analytic approximation fits well at low energies.  The analytic approximation
begins to diverge from the numerical result for $\gamma \gtrsim 1.3$, when the dark matter contribution
to the potential begins to dominate at small $r$.  Note, however, that for smaller values of $\rho_s$,
the power-law result becomes a better approximation to the full numerical result, even for larger $\gamma$.

But even if the baryonic potential dominates, the linear
approximation to this potential is only very good well
within the bulge; at the edge of the bulge, the gravitational
potential will necessarily deviate from the linear approximation,
causing $P_n^2$ to deviate from power-law form at the edge of the
bulge.  This can cause the angular distribution to deviate from
the analytic power-law form even at small angle, because the
analytic approximation mismodels the velocity-distribution at
the edge of the bulge, even along a line of sight aimed directly
at the GC.  The accuracy of the analytic approximation to the
angular distribution thus depends on the steepness of the profile;
for a steeper profile, with a larger fraction of the total annihilation rate concentrated deep in the interior of the bulge,
the analytic result will be better.  The larger the velocity power-law
exponent $n$, the steeper the profile must be in order
for the analytic approximation to be valid, since a larger value of
$n$ implies that dark matter annihilation is increasingly
dominated by high-speed particles which explore the edges of the
Bulge.

We thus see that changes to the steepness of the density
slope within the bulge produce two competing effects.  A
steeper profile tends to concentrate dark matter annihilation
deep with in the bulge, making the analytic result a better
approximation.  More generally, for a steeper profile, dark matter
annihilation in the bulge is dominated by particles which
never leave the bulge, implying that the angular distribution is
determined only by the potential within the bulge, and is decoupled
from what goes on outside.  For a more shallow cusp, one must
understand the details of how gravitational potential changes as
one approaches the edge of the bulge.  In this case, one would
similarly expect that the effects of triaxiality, as well as
deviations from the spherical approximation to the baryonic potential (due, for example
to the Galactic disk contribution), will become more important.  For larger $n$, one requires a
steeper profile in order to decouple these effects, since
the annihilation rate is increasingly dominated by
high-speed particles which, though a small fraction of the
dark matter within the bulge, nevertheless have an enhanced
annihilation rate.

But the steeper the dark matter density slope, the more dark matter
will tend to dominate the gravitational potential deep within the
bulge, which would also invalidate the analytic approximation.  The strength
of the dark matter gravitational potential depends on both $\gamma$
and $\rho_s$.  If one assumes a generalized NFW profile throughout the
entire MW halo, then one can estimate $\rho_s$, with uncertainties, from
observational data throughout the halo.  But if the the dark matter distribution
within the bulge forms a separate power-law distribution, then the normalization
of the density distribution within the bulge is much less constrained, since it
need only smoothly match onto a shallower generalized NFW outside the bulge.

\begin{figure}[p!]
    \centering
    \includegraphics[width=0.5\textwidth]{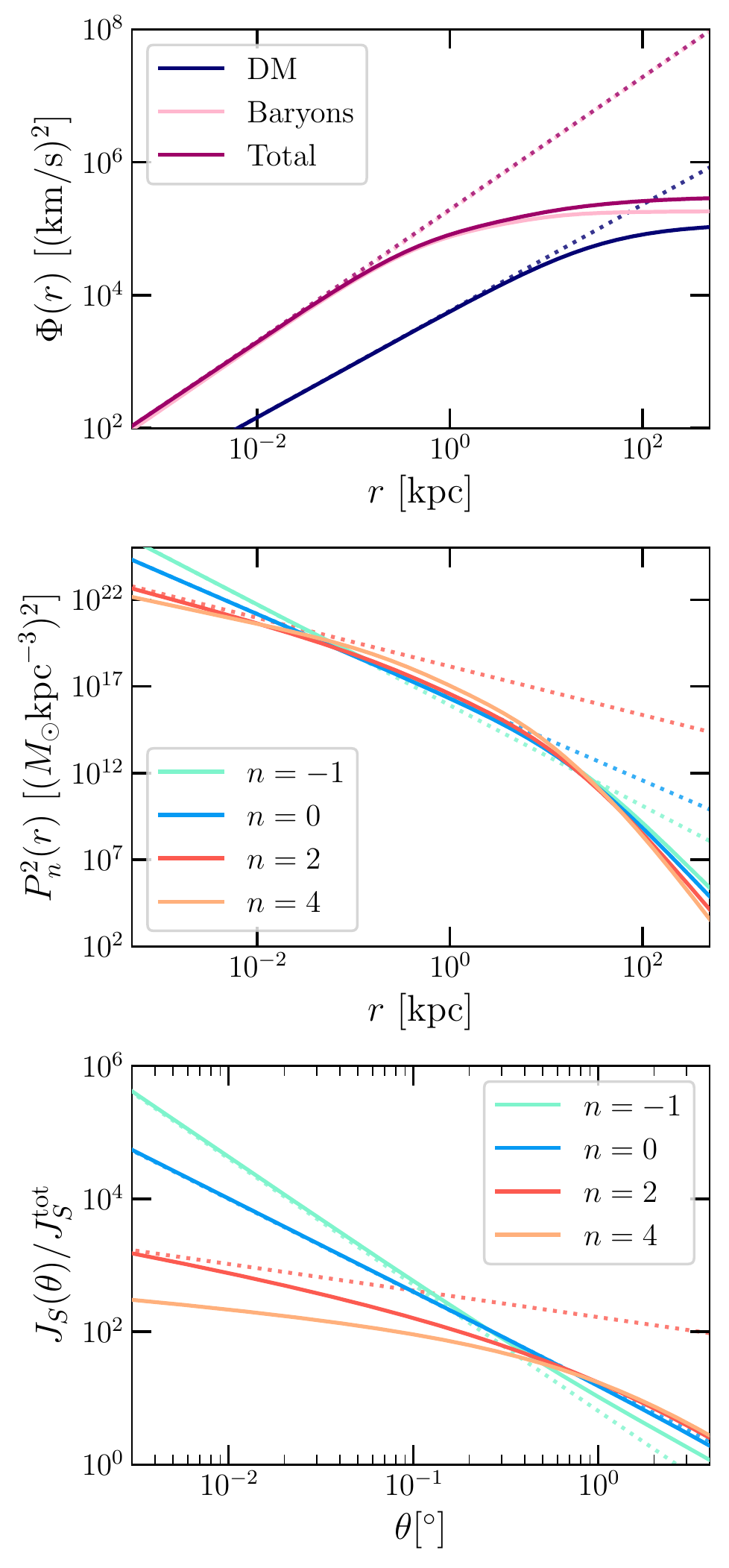}
    \caption{The top panel shows the gravitational potential $\Phi$ for the DM contribution, baryonic contribution, and the total potential for an NFW distribution with $\gamma=1.2$ and $\rho_s=4\times 10^6 M_\odot / \kpc^3$. In all three panesl, the dotted lines show the analytic approximation for comparison.
    The middle panel shows $P^2_n(r)$ for all $n$ considered. The bottom panel shows the $J$-factor distributions $J_S(\theta)$ for all $n$ considered normalized by $J_S^{\rm tot}\equiv \int_0^{4\pi/180}d\theta\sin\theta J_S(\cos\theta)$. The analytic comparisons for the normalized J-factors are matched to the numerical calculations at $\theta=1\times10^{-3}$ degrees for ease of comparison.}
    \label{fig:analytic_approx}
\end{figure}

As an illustrative example, we consider the case $\gamma = 1.2$, with
$\rho_s = 4 \times 10^6 M_\odot / \kpc^3$.  For this choice, the dark matter density at the solar radius
would be give by $\rho_\odot = 0.25~\gev / \cm^3$.  In Figure~\ref{fig:analytic_approx}, top panel,
we plot the potential $\Phi(r)$,
including the
baryonic and dark matter contributions, as well as the total potential.
The solid lines are the full potential, while the dashed lines represent the analytic
power-law approximation which we consider.  We see that, for this choice of $\gamma$ and
$\rho_s$, the baryonic contribution  dominates the total potential, which is required for
the analytic approximation to be valid.  But the power-law approximation to the baryonic potential
breaks down once we reach the edge of the bulge, for $r \gtrsim {\cal O}(0.1)~\kpc$.  As such, we expect deviations
from the analytic approximation to be driven by particles which explore the edge of the bulge.

In the middle panel of Figure~\ref{fig:analytic_approx}, we plot $P_n^2$ for $n=-1, 0, 2$ and $4$.  Again, the complete numerical
calculation is shown in solid lines, while the power law analytic approximation is shown in dashed lines.
For $n=0$, the ordinary case of $s$-wave annihilation, the analytic approximation is nearly exact
within the bulge, since dark matter annihilation is velocity-independent, so deviations from the
analytic approximation to the potential are irrelevant.  For the case of Sommerfeld-enhanced
annihilation ($n=-1$), the analytic approximation
to $P_{n=-1}^2$ matches the numerical calculation well inside the bulge, because for Sommerfeld-enhanced
annihilation, the annihilation rate is dominated by low-speed particles which never explore the edges of
the bulge, where the potential deviates from power-law.  For $p$-wave annihilation ($n=2$), we see that
the power law approximation is only a good fit deep within the bulge.  This is not surprising, since the profile
we have chosen is only slightly steeper than the limit $\gamma =1$, at which the analytic approximation breaks
down entirely for $p$-wave annihilation, which is then dominated by particles which leave the bulge.

In the bottom panel of Figure~\ref{fig:analytic_approx}, we plot the angular distribution, which
is proportional to $J_S (\cos \theta)$, for $n=-1, 0, 2$ and $4$.  The angular distribution is normalized
to unity when integrated over $4^\circ$,
and is plotted with solid lines.
We used dashed lines to plot the power-law analytic approximation to $J_S (\cos \theta)$, which is normalized
(for ease in comparing to the numerical result) to match the numerical computation as $\theta \rightarrow 0$.
As expected from the discussion of $P_n^2$, the slope of analytic approximation matches the that of the
numerical computation fairly well within the inner degree, for Sommerfeld-enhanced annihilation.
On the other hand, for the case of $p$-wave annihilation, the angular distribution matches the analytic
prediction only for a very small angular range ($\Delta \theta \sim {\cal O}(10^{-3})$ degrees), which would not
be useful for a data analysis.

It is interesting to note that, for this case ($\gamma = 1.2$), the photon angular distribution arising from
Sommerfeld-enhanced annihilation is a well-matched by the analytic approximation to the GC because dark matter
annihilation is dominated by low-speed particles which only explore the bulge.  This result thus largely
depends only on the slope of the dark matter distribution within the bulge, and is independent of the dark matter
distribution outside the bulge.  This analytic approximation can thus be generalized beyond the assumption of an
NFW profile, and does not assume that the slope of the profile inside the bulge is that same as that outside.  Similarly,
it is relatively robust against the effects of triaxiality, which is likely to affect the dark matter profile
at relatively large distances from the GC.

On the other hand, the analytic approximation to $p$-wave fails (except at very small angles) because dark matter
annihilation is, in this case, dominated by high-speed particles which can leave the bulge.  Interestingly, this is the
case even though, for $\gamma = 1.2$, the analytic approximation to the velocity distribution matched the numerical
calculation fairly well (see Figure~\ref{fig:fe}).  Beyond the failure of
the analytic approximation to the angular distribution, the more general lesson is that, for $p$-wave annihilation
in the case of a profile which is not very steep, the angular distribution near the GC cannot be determined accurately
without a full knowledge of the dark matter profile and baryonic potential, even far away from the GC.  Similarly, the
effects of triaxiality, which are expected to be relatively small within the Galactic bulge, should be expected to nevertheless
have a significant effect on the angular distribution at small angles.  Even though a
negligible fraction of the DM within the Galactic bulge may explore the gravitational potential very far away, that small
fraction of particles will dominate the annihilation rate (and thus the angular distribution) unless the profile is very steep.
The considerations are even more relevant for the case of $d$-wave annihilation.

\section{The Galactic Center Excess}
\label{sec:GCE}

As an application, we consider the GC excess.
The excess arises in photons with an energy of a few $\gev$,
at which the Fermi-LAT would have an angular resolution of
order a few tenths of a degree.  But for the purpose of understanding
how well the analytic approximation works, and when and why it fails,
we will consider angles as small as $10^{-3}$ degrees from the GC.

For the
case of $s$-wave dark matter annihilation ($n=0$), the
angular distribution of the GC excess would require an
inner slope within the bulge of $\gamma = 1.2-1.4$~\cite{Hooper:2010mq,Calore:2014xka}.  For $p$-wave annihilation
($n=2$), the same morphology would require $\gamma = 1.7 - 1.9$, if the analytic approximation is valid.
This range of $\gamma$ is steep enough that dark matter annihilation
deep within the bulge would be dominated
by particles which never left the bulge.
Note that, although this is a much steeper profile
than is usually considered, the slope may be much shallower
outside the bulge.
A model of $d$-wave annihilation would also match the
angular distribution of the GC excess if $\gamma = 2.2-2.4$.
We will focus on the case of $p$-wave annihilation, with
$\gamma = 1.7$.

But for such a steep profile the baryonic contribution
to the gravitational potential may no be longer dominant.  This
may be the case, but need not be.
One can always reduce the amplitude of the dark matter density
within the bulge, in order to ensure that the baryonic
potential dominates, with the amplitude of the GC excess
obtained by a corresponding rescaling of $(\sigma v)_0$.  However,
if the dark matter density at the edge of the bulge is too small,
it would be difficult to obtain a density at the solar radius which
is consistent with observational data.  If the dark matter density at
the solar radius is $0.3~\gev / \cm^3$, and if the profile outside the
bulge is NFW with $r_s = 21~\kpc$, then the dark matter density at
$r=c_0$ would be $\rho_{c_0} = 2 \times 10^8 M_\odot / \kpc^3$.  Setting
$\rho_{c_0} = \rho_s (c_0 / r_s)^{-1.7}$, we would
find $\rho_s = 5 \times 10^5 M_\odot / \kpc^3$.  If we had instead assumed that the
dark matter density between the edge of the bulge and scale radius had
a slope of only $-0.6$, this would have reduced $\rho_{c_0}$, and in turn  $\rho_s$, by a factor of 2.5.
We will consider the cases $\rho_s = 4 \times 10^a M_\odot / \kpc^3$, with
$a=3, 4, 5, 6$.

In determining the angular distribution with a full numerical calculation,
it is necessary to make some assumption for the profile outside the bulge.
For simplicity, we will assume a generalized NFW profile with inner slope
of $\gamma = 1.7$ throughout the MW halo.  But as we have noted, such a
steep profile may not be valid outside the bulge.  To characterize
the extent to which DM annihilation outside the bulge affects the angular
distribution, we will perform the numerical calculation in two ways.  First,
we compute the complete angular distribution assuming a generalized NFW profile.
Second, we will compute the velocity-distribution assuming a generalized NFW profile
throughout the MW halo, but will compute the $J$-factor by assuming that there is
no dark matter annihilation outside the Bulge.  This amounts to taking the upper
limit of integration in eq.~\ref{eq:JS_exact} to be $c_0$.

\begin{figure}
    \centering
    \includegraphics{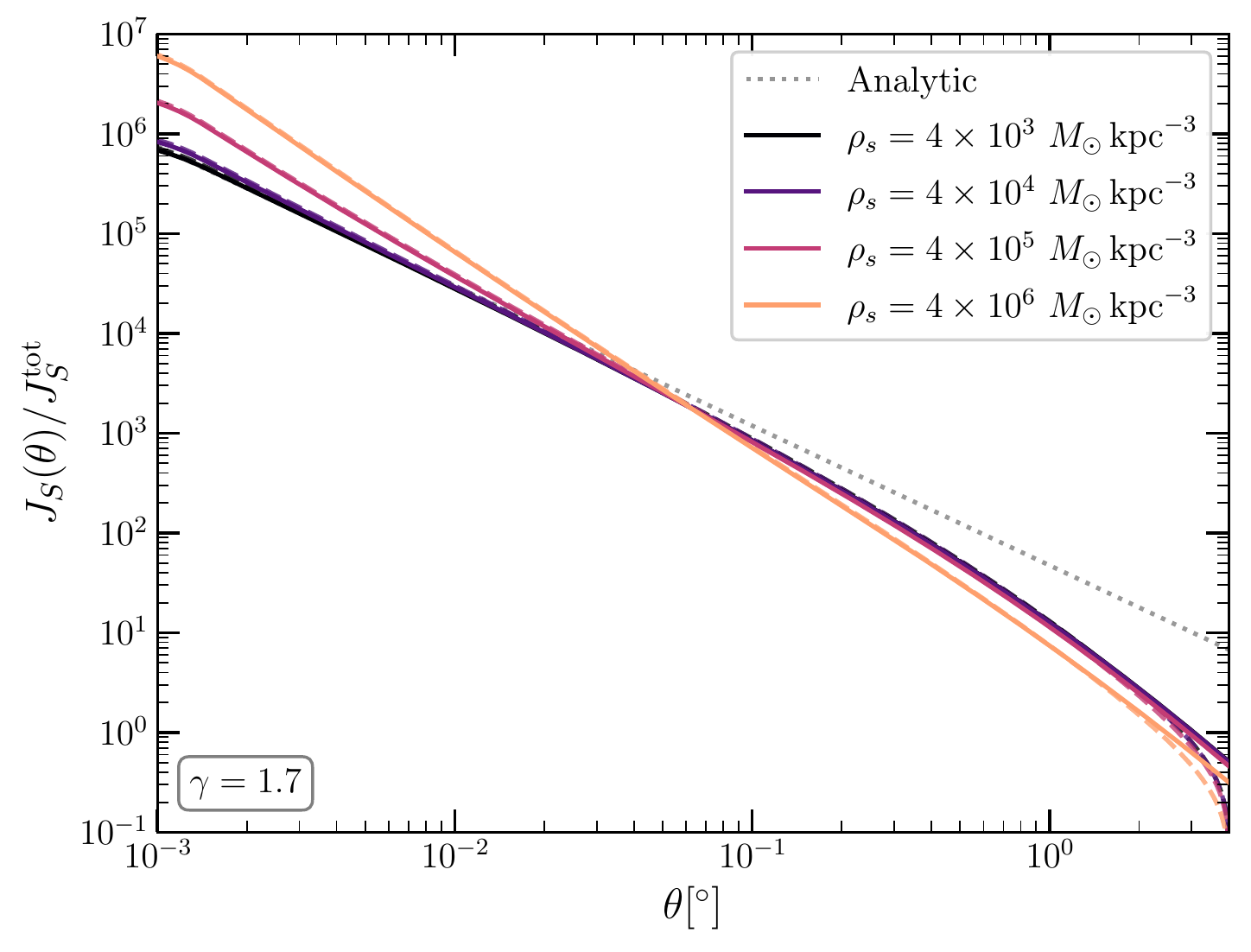}
    \caption{$J$-factors for varying $\rho_s$ values for an NFW distribution with $\gamma=1.7$ normalized by $J_S^{\rm tot}\equiv \int_0^{4\pi/180}d\theta\sin\theta J_S(\cos\theta)$. The numerical calculations with the DM annihilation truncated outside the bulge are shown as dashed lines. The analytic approximations are matched to the numerical calculations at $\theta=1\times10^{-3}$ degrees for ease of comparison.}
    \label{fig:gamma17}
\end{figure}

We plot our results in Figure~\ref{fig:gamma17}.
The solid lines indicate the numerically-computed angular distribution for the
case of $p$-wave annihilation,
normalized so that the integral to $4^\circ$ is unity.
The dashed lines are similar, but neglecting dark matter annihilation outside
of the core.  Finally, the dotted line is the analytic approximation, which is
normalized to match the numerical calculation for $\rho_s = 4 \times 10^3 M_\odot / \kpc^3$
as $\theta \rightarrow 0$.  Note that the solid and dashed lines are nearly identical.  This
implies that, at small angles, the dark matter annihilation rate is indeed dominated by particles which
annihilate within the bugle itself.

We see that, as $\rho_s$ decreases, the analytic approximation becomes a better match to
the numerical calculation, for a larger range of angles.  This is to be expected, because, this
limit amounts to a best case scenario, where the profile is steep enough that even the high-speed
particles which dominate $p$-wave annihilation are less likely to explore the edge of the bulge.  But
the dark matter density within the bulge is also taken to be small enough that the dark matter does not
deform the potential significantly.  But even in this case, we see that the analytic approximation
begins to diverge from the numerical calculation for angles of ${\cal O}(0.1^\circ)$.  This essentially happens
because, no matter how steep the profile, the effect can only be to concentrate particles near
the GC.  To the extent that the baryonic potential dominates the profile, there will always be a very
small high-speed tail of particles which can reach the edge of the bulge, but this small tail will
nevertheless give a large contribution to the annihilation rate for the case of
$p$-wave annihilation.

We may thus draw a broader lesson from the comparison of the analytic approximation to the
numerical calculation of the angular distribution in the limit of  small $\rho_s$.  The difference between
these curves roughly characterizes the dependence of the angular distribution, even well within the
bulge, on particles which leave the bulge, and thus the level of uncertainty in the angular distribution
introduced by variations in the dark matter profile form, triaxiality, or any other features of the
dark matter or baryonic distribution outside the bulge.

It is also interesting to note that, as $\rho_s$ increases, the steepness of the angular
distribution at small angles also increases.  This result can also be understood intuitively.
As $\rho_s$ increases, the dark matter contribution to the gravitational potential becomes
more important.  In the limit in which the dark matter contribution to the potential dominates,
we may ignore the baryons entirely.  This problem was considered in~\cite{Boucher:2021mii}, where it
was found that the angular distribution (at small angle) for $p$-wave annihilation is a power law with slope
$3-3\gamma$.  We would thus expect the slope of the angular distribution to increase by a factor of $3/2$, as $\rho_s$ increases.
This result is confirmed in Figure~~\ref{fig:gamma17}, where the choice
$\rho_s = 4 \times 10^6 M_\odot / \kpc^3$ leads to an angular distribution with a slope of $\sim -2.1$ at
small angle.

Interestingly, if the angular distribution is to have slope $-1.4$ at small angle
(as one would expect for $s$-wave annihilation with $\gamma = 1.2$), then for $p$-wave annihilation in a
dark matter-dominated halo, one would require $\gamma \sim 1.47$.  Out to distances of a few $\kpc$, for which
the small angle approximation is still valid, the gravitational potential varies between being either baryon-dominated
or dark matter-dominated,  If $p$-wave annihilation were to produce an angular distribution consistent with what is observed for
the GC excess, one would expect the slope of the dark matter density profile to lie in the $\gamma \sim 1.5 - 1.7$ range.

Note that, in the case of $p$- or $d$-wave annihilation,
a steeper slope within the inner slope region could also
lead to a larger rate for dark matter annihilation near the
black hole at the center of the MW~\cite{Shelton:2015aqa,Sandick:2016zeg}.
We have not included the effects of this on our analysis,
but this would be an interesting topic for future work.

\section{Conclusion}
\label{sec:conclusion}

We have considered velocity-dependent dark matter annihilation within the
Galactic bulge. Because the rate of velocity-dependent annihilation at any
given location depends not only on the dark matter density at that point,
but on the gravitational potential at all locations sampled by particles passing
through that point, determining the annihilation rate can be a very non-local
problem.  Our goal has been to understand the extent to which the angular distribution
of photons arriving from the direction of the bulge can understood entirely using
features of the dark mater and baryonic density distributions within the bulge.  Because the behavior
of the gravitational potential within the bulge is very different from its behavior far away,
the behavior of the dark matter density profile may also be quite different from
what is expected from simulations which are extrapolated to small distances.  There are large uncertainties in our
ability to probe the dark matter profile in baryon-rich environments such as the bulge,
either using simulations or stellar tracers, making it important to understand how strongly
these environments control the angular distribution.

The GeV excess of photons from the GC  may have its origin in dark matter annihilation.
But such solutions are constrained by searches for photons from other dark matter-rich
environments, such as dSphs.  Scenarios of velocity-dependent dark matter annihilation
can avoid such constraints, which makes it particularly interesting to determine not
only if these scenarios can reproduce the observed angular distribution, but also to
determine which features of distribution contribute to this determination.

We have found that for the case of Sommerfeld-enhanced annihilation, dark matter annihilation
within the bulge is typically dominated by slow-moving particles which never leave the bulge.  In this
case, the photon angular distribution at small angle is largely controlled by a single parameter:
the dark matter density slope within the bulge.  The behavior of the dark matter distribution
outside the bulge, including the slope in the region between the bulge and the scale radius,
has only a small effect.  In this case, the photon angular distribution largely probes the
localized astrophysics of dark matter within the bulge, which can be robustly reconstructed
if the angular resolution (for the energy range of the photons produced by DM annihilation) is
$\lesssim 0.1^\circ$.

On the other hand, for $p$- or $d$-wave annihilation, dark matter annihilation within the
bulge receives a significant contribution from high-speed particles which leave the bulge.
Although this is a small fraction of the dark matter particles within the bulge, this energetic tail
nevertheless dominates the annihilation rate for these particular scenarios of velocity-dependent
annihilation.  In this case, the photon angular distribution, even at small angle, necessarily
depends on the dark matter profile and the gravitational potential well outside the bulge.  As a
result, uncertainties in the dark matter profile, including the effects of triaxiality and the way
the dark matter density interpolates between its behavior within and outside the bulge, will
necessarily have a non-trivial impact on the angular distribution.

The GC excess, as currently understood, would be consistent with $s$-wave dark matter annihilation
with a density slope of $\gamma = 1.2-1.4$.  To alleviate constraints from dSphs searches, one would
like to consider a scenario of $p$-wave annihilation.  Our results have shown that, in this case,
it is not possible to cleanly relate this angular distribution to the slope of the density profile.
Instead, we find a general prediction that this angular distribution would require a steeper
profile in the case of $p$-wave annihilation.  For example, the angular distribution (at small angle)
yielded by $s$-wave annihilation with $\gamma=1.2$ inside the bulge, would be yielded with $p$-wave
annihilation with $\gamma \sim 1.5 - 1.7$, with the details determined by parameters such as the
ratio of baryons to dark matter in the MW, the slope of the dark matter distribution outside the bulge,
triaxiality, etc.

Note, we have not attempted an actual fit to data from the GC.  Instead, we have simply taken at
face value the detailed analyses performed in previous works, which have considered Fermi data from
the GC as well as a variety of backgrounds in order to assess the morphology of the excess.
But millisecond pulsars may generate part or all of the GC excess.  Even assuming a large contribution
to the excess arises from DM annihilation, correct inclusion of a MSP contribution could change the
morphology of the contribution arising form DM annihilation.  Similarly, improvements in our understanding
of other backgrounds could also modify our understanding of the excess morphology.    Although this
would modify the details of our analysis, the overall framework would remain unchanged.  A more detailed
study of $p$-wave annihilation as an explanation for the GC excess seen in Fermi data would be an
interesting topic of future work.

{\bf Acknowledgements}

We are grateful to Pearl Sandick and Louis E.~Strigari for useful discussions.
JK is supported in part by DOE grant DE-SC0010504.
JR is supported by NSF grant AST-1934744.

\bibliography{thebibliography.bib}
\end{document}